# A new distance law of planets and satellites in the solar system

Cho Byong Lae[1], Kim Jik Su[2], Ri Yong Il[2], Kim Chol Jun[3], Pak Yong Chol[2], Kim Ki Chol[2], Mun Ui Ri[3]

[1]Department of Theoretical Physics, Institute of Physics, the State Academy of Sciences, DPR Korea.
[2]Department of Astrophysics, Pyongyang Astronomical Observatory, the State Academy of Sciences, DPR Korea.
[3]Faculty of Physics, **Kim Il Sung** Universiry, DPR Korea.

**Abstract**

In the 60's of last century, it has been substantiated that an equation of Schrödinger type could describe the diffusion phenomena, and the main consequence from this finding has been that there would be wave property in the diffusion processes as well. This theory has been immediately proved through laboratorial experiments.

Afterwards the theory was applied to the primordial nebula which was thought to surround the protosun, and has found the consistency of the prediction of the theory with current distance distribution of the planets to be excellent.

At the end of 20[th] century new satellites of planets were discovered. On the basis of the new data, the theory is tested thoroughly and the result allows us to come to the conclusion that the basic process for the distances of the planets from the protosun to be determined has been the diffusion of the primordial nebula consisting of mainly molecular gas.

1. Introduction

Up to the 60's of the 20[th] century, equation describing the diffusion process only has been the Fokker-Planck equation along with usual diffusion equation. In the mid-twentieth century in Ref. [1], the author first shown that an equation of Schrödinger type could be derived from the Kolmogorov equation describing Markov chain as a stochastic process. The Fokker-Planck equation describes the diffusion of gas through the probability density $\omega$ while the author of Ref. [1] represented the density $\omega$ as $\omega = \psi\psi^*$ by introducing the state function $\psi$. Then the Fokker-Planck equation goes over to an equation of Schrödinger type, which is expected to describe the diffusion as well if the Planck constant $\hbar$ is replaced by $D = \dfrac{\hbar}{2m}$ where $m$ is mass of a diffusing particle and $D$ is diffusion coefficient. In such a way, the author of Ref. [1] showed that the usual diffusion phenomena could be described also by an equation mathematically quite equivalent to the Schrödinger equation when one defined the state function $\psi$ of a diffusing particle. The author of Ref. [1], however, only argued the problem from the point of view of a possibility of causal interpretation of the quantum mechanics, and had not considered the applicability of the Schrödinger type equation to actual diffusion processes.

In the 60's of the 20th century, the author of Ref. [2] had first applied the Schrödinger type equation to the diffusion phenomena and had first investigated the wave character manifested in the diffusion process. The Schrödinger equation, as is well-known, describes the wave-particle duality inherent in quantum system, so it is natural to expect some manifestation of the wave property in a system lying in a state of pure diffusion. Ref. [6, 7] derived this equation from the hydrodynamic point of view applying the variation method and has considered more vividly its physical significance in the diffusion processes. On the basis of this theory, Ref. [4] expounded the physical essence of the Riesegang phenomenon that was known in physical chemistry from the 19th century. The authors of Ref. [3] have performed an interesting experiment according to which when a solution of copper sulphate was electrolyzed, the plus ions of copper passed through a small hole of screen dipped into the solution between copper and zinc plates formed a well-known diffraction pattern on the zinc plate. That diffraction pattern could be excellently explained by the wave property inherent in the diffusion process of copper ions.

Ref. [2, 5, 6, 7] furthermore applied the theory to the primordial nebula of the solar system and has derived a new distance law of the planets which was consistent with observation. It reads

$$\sqrt{R_m} = K\,(m+1),\ (m = 1,\ 2,\ 3,...), \tag{1}$$

where $K$ is a constant depending on the temperature, density of diffusing medium and mass of central protostar. The author of Ref. [5] divided the planets of the solar system into inner and outer parts, and estimated the constant $K$ from the observation, according to which $K_{in} = 0.21$ for the inner group of planets and $K_{out} = 1.08$ for the outer group (see Table. I).

Ref. [2, 5, 7] further showed that the distance law (1) was consistent with observation for the satellites of the planets as well, and has estimated the constant $K$ for the every system of these satellites. The author of Ref. [6] has made an attempt to explain the apparent band pattern of the head of comet by a manifestation of the wave phenomenon in diffusing gas of heated head of the comet approaching to the Sun.

To summarize, the fact that the law (1) on the regularity in distance distribution of planets and satellites as a manifestation of the wave property occurring in the diffusion process is consistent with observation suggests the necessity to regard the diffusion process having taken place in the primordial gaseous nebula as essential in the formation of planets and satellites.

An earliest empirical rule of Titius-Bode, $R_n = 0.4 + 0.3 \times 2^n\ (n = 0,1,2,\cdots)$ is compatible with observation only for the planets from Venus (n=0) to Uranus (n=6) and Mercury is to be assigned to $n = -\infty$, while for Neptune (n=7) the rule yields a distance 1.3 times as large as the actual distance. In the case of Pluto, the consistency is far worse.

In the 40's of the 20th century, O.Yu.Shmidt has explored the formation of the planets in the solar system in terms of a capture of gas-dust cloud and has obtained a formula for distance distribution of the planets that was similar to the formula (1) [8,9]. It reads

$$\sqrt{R_n} = a + b \cdot n, \qquad (n = 1, 2, 3, \ldots) \tag{2}$$

Where $a$ and $b$ are constants which have to be determined from the observation. The similarity of (1) and (2) would have the constant $a$ and $b$ give a definite significance, but the physical mechanisms governing the formulas are quite different, so that the comparison of the formula (2) with the formula (1) would be meaningless.

The proto-planet theory [15, 16], capture theory [17], the solar nebula theory [12, 18] and the modern Laplacian theory [19] which are published after 1960s, have succeeded to explain many important features observed in the solar system. However, they have not given the explanation for distance distribution of the planets and the satellites from their central bodies.

The reference [13] investigated the mass distribution of the solar system from a point of view of pure statistical mechanics and compared it with actual smoothed mass distribution of the planets.

In this work, following the previous approach of Ref. [2, 6, 7] we will testify the distance law (1) more extensively making use of the improved data of distance of the planets and new discovered satellites of the solar system.

This article is organized as follows. In Sect.2, the basic theory regarding the statistical kinetic equation and its physical significance represented in the manifestation of the wave property in diffusion phenomena are described. In Sect.3 the distance distribution law of the planets and the satellites of the solar system is derived in the most general form. In Sect.4 the actual distances of these planets are compared with the prediction of the theory. In Sect.5, for the satellites of every planet, from Jupiter to Neptune, distances from a central planet are compared with the theory and the constant $K$ are derived therefrom. In Sect.6, the results are summarized and the general implication of the results is discussed.

2. **Statistical kinetic equation of diffusing particles and wave property manifested in diffusion**

Ref. [1] has first derived the Schrödinger type equation describing the diffusion phenomena on the basis of the mathematical formalism of the Markov chain as a stochastic process. Ref. [2] has investigated the equation and derived it from the variation principle. Ref. [2, 3, 4] has first applied the theory to clarify the wave property manifested in the diffusion process. In this section, we review the mathematical formalism of the theory and consider the physical implication of the wave character inherent in diffusion phenomenon.

The statistical kinetic equation derived in Ref. [2] reads as follows

$$2iD\frac{\partial \psi}{\partial t} = -2D^2 \nabla^2 \psi + U\psi ,\qquad(3)$$

where $\psi(x,t)$ is called a state function of diffusing particle and $D$ is the diffusion coefficient. This equation bears a resemblance exactly to the Schrödinger equation if one puts.

$$D = \frac{d}{2\mu} .\qquad(4)$$

We obtain the Schrödinger equation.

$$id\frac{\partial \psi}{\partial t} = -\frac{d^2}{2\mu}\nabla^2 \psi + V\psi ,\qquad(5)$$

where

$$V = \mu U \qquad(6)$$

The putting $d \to \hbar$ yields just the Schrödinger equation.

The Schrödinger equation implies all the wave characteristics of the microprocess in quantum level.

It is natural to expect the equation (3) mathematically quite equivalent to the Schrödinger equation to imply some wave property in the diffusion phenomena as well. The equation (3), however, is never physically equivalent to the Schrödinger equation because although the dimension of $d$ is the same as one of the Planck constant $\hbar$, but the value of quantity $d$ is about $10^{-7} gcm^2/s$ in temperature $T \approx 300K$ and density $N \approx 10^{13} cm^{-3}$ while the Planck constant $\hbar$ is $10^{-27} gcm^2/s$ in order. Namely, the quantity $d$ is about $10^{10}$ times as large as the Planck constant. This means that the diffusion equation (3) describes macroscopic processes of the diffusion in contrast to the microscopic quantum processes.

The state function $\psi(x,t)$ of a diffusing particle represents statistically the probability of diffusing particles through the relation $\psi^*\psi = \rho$, where $\rho$ is the probability density of diffusing particles. The asterisk * denotes complex conjugate.

In the case that the potential V does not depend on time the state function $\psi(x,t)$ can be expressed as $\psi(x,t) = \Phi(x)e^{-\frac{iE}{d}t}$ where $E$ is a constant and $\Phi$ is a function of position of a particle. Then the function $\Phi(x)$ satisfies an equation of stationary state

$$-2D^2\nabla^2\Phi + U\Phi = E\Phi/\mu. \qquad (7)$$

This is just the statistical kinetic equation for the stationary state and the original equation (1) refers to as statistical kinetic equation of diffusing gas. This equation only has solutions in definite discrete quantities of $E$ which are eigenvalues of the equation (7), and corresponding solutions are eigenfunctions. The physical significance of the quantity $E$ may be interpreted as mean energy of a diffusing particle.

The wave property exhibited in the diffusion has been demonstrated in an experiment where the diffraction pattern has observed in electrolysis [3]. Between two electrolytic plates one sets up a screen where there is a small hole. When the current flows between two plates the current is possible to only pass through the hole on the screen, and the current which constitutes of plus ions exhibits some interference pattern on the minus plate, which bears a resemblance to the diffraction pattern of light passed through a small hole. The experiment has shown that this diffraction image could not be interpreted by de Broglie wave and could be expounded only in terms of the macroscopic diffusion described by the statistical kinetic equation of the diffusing particles (7) [3]. This experiment has given a crucial evidence for the existence of the wave character in the diffusion.

The wave vector in this diffusion phenomenon can be written as

$$\mathbf{k} = \frac{p}{d} = \frac{\mu\upsilon}{d} = \frac{\upsilon}{2D} \qquad (8)$$

where the relation (4) is used, p is a momentum of the wave and d is a constant introduced in the relation (4). In the kinetic equation (5) the constant d replaced the Planck constant $\hbar$, so de Broglie equation p=$\hbar$k is possible to be replaced by p=$d$**k** and the relation (8) holds. The wavelength is therefore,

$$\lambda = \frac{2\pi}{k} = \frac{4\pi}{\upsilon}D \tag{9}$$

The drift velocity of plus ions $\upsilon$ may be expressed by intensity of the electric field between two plates $\varepsilon$ and mobility $\alpha$ of plus ions as follows.

$$\upsilon = \alpha\varepsilon \tag{10}$$

On the other hand, the diffusion coefficient $D$ and the mobility $\alpha$ are related by an equation (Einstein relation).

$$\frac{D}{\alpha} = \frac{k_B T}{e}, \tag{11}$$

where $k_B$ is Boltzman constant, $e$ charge of ion ($Cu^{++}$) and $T$ temperature of the electrolyte. Putting (10) and (11) into relation (9) one obtains a relation

$$\lambda = \frac{4\pi k_B T}{e\varepsilon}, \tag{12}$$

This makes the wavelength of the diffusion wave evaluate through the temperature and the intensity of electric field. The validity of the relation (12) has been demonstrated in the above-mentioned interference experiment.

The well-known relation on the position of rings of the maximum intensity in the diffraction image,

$$\sin\varphi_m = k_m \frac{\lambda}{a} \tag{13}$$

may be used where $\varphi_m$ is angle of the position of maximum intensity, $k_m$ a constant depending on the order of the maximum and $a$ size of the hole. Putting the relation (12) into (13) one obtains a final relation

$$\sin\varphi_m = \frac{k_m}{a}\frac{4\pi k_B T}{e\varepsilon}. \tag{14}$$

This formula shows that if the pattern appeared on the minus electrolytic plate is interference rings due to the wave property implied in the diffusion process the positions of maximum in the pattern have to depend on the temperature of the electrolyte and electric field between two plates. Actually, the experiment has shown exactly the dependence of wavelength on the temperature and electric field. Thus the existence of the wave property in the diffusion has been proved experimentally in the 60's of the last century.

## 3. Derivation of the distance law

Suppose that the primordial solar nebula has consisted of almost homogeneous gas. We suppose also that first appearance of the inhomogeneity in the protonebula was due to the wave property of the diffusion rather than the gravitational instability. In center of the nebula the protosun has been forming and the thermonuclear reaction has not been yet initiated, so the surrounding protonebula has been exposed to the sunlight, and the inner part of the nebula might be heated until about a few thousands degree. Molecules of the inner part of the protonebula would be dissociated into individual atoms whereas the outer part of the nebula would have a temperature of about a hundred degree or less. Inner and outer parts of the protonebula, on account of the difference in temperature, density and constituents, would have gone under different diffusion circumstances.

In general, the protonebula surrounding the protosun has been under gravitational field of the Sun, and the potential $U$ in equation (7) is $-\frac{GM_\odot}{r}$, and in terms of the relation, $D = d/2\mu$, the equation (7) may be written as follows

$$-\frac{d^2}{2\mu}\Delta\Phi - \frac{GM_\odot \mu}{r} = E\Phi, \qquad (15)$$

where $\Delta$ is Laplasian, $\mu$ and $M_\odot$ masses of a diffusing particle and the Sun, respectively, $G$ Newton's gravitation constant, $r$ distance from the Sun and $E$ total energy of diffusing particle. The equation (15) resembles the Schrödinger equation of stationary state for an electron revolving round the nucleus of hydrogen

$$-\frac{\hbar^2}{2\mu}\Delta\psi - \frac{e^2}{r}\psi = E\psi. \qquad (16)$$

Therefore, without solving the equation (15) one can find its eigenvalues making use of the conventional solution of the equation (16).

As is well known, the eigenvalues of the equation (16) are represented by

$$E_n = -\frac{e^4 \mu}{2\hbar^2}\frac{1}{n^2}. \quad (n = 1,2,3\cdots) \qquad (17)$$

Hence, the energy eigenvalues of the equation (15) may be obtained by replacing

$\hbar \to d = 2\mu D$ and $e^2 \to GM_\odot \mu$ as follows

$$E_n = -\frac{G^2 M_\odot^2 \mu}{8D^2}\frac{1}{n^2}. \tag{18}$$

Orbital angular momentum of a diffusing particle and its component are expressed as in the case of hydrogen atom as follows

$$L_l = d\sqrt{l(l+1)} = 2\mu D\sqrt{l(l+1)}, \quad (l = 0,1,2,3,\cdots), \tag{19}$$

$$L_m = md = 2\mu Dm, \quad (m = 0, \pm 1, \pm 2 \cdots \pm l). \tag{20}$$

Express the eigenfunctions corresponding to the eigenvalues $E_n$ of equation (15) as $\Phi_n(r) = \Phi_n(r,\vartheta,\varphi)$ where $r, \vartheta$ and $\varphi$ are spherical coordinates, $z$ coordinate being parallel to the direction of the angular momentum. Then $|\Phi_n(r)|^2$ is the probability density for a diffusing particle to be found in a position $r$ from the Sun. Product of $|\Phi_n(r)|^2$ and total number $\tilde{N}$ of particles surrounding the Sun yields the spatial density of gaseous protonebula in a position $r$. This spatial distribution density will give a number of gaseous bands distinguished by the number $n$. In order to find the distribution law of the bands depending on the distance $r$ from the center of the nebula one can use the virial theorem as in the quantum mechanics. It reads

$$\langle H \rangle_n = \frac{1}{2}\langle V \rangle_n. \tag{21}$$

In this relation $\langle H \rangle_n$ and $\langle V \rangle_n$ represent total average energy and potential energy of a particle, respectively, as follows

$$\langle H \rangle_n = \int \Phi_n^* \hat{H} \Phi_n d\tau = E_n = -\frac{G^2 M_\odot \mu}{8D^2}\frac{1}{n^2}, \tag{22}$$

$$\langle V \rangle_n = -\int \Phi_n^* \frac{GM_\odot \mu}{r} \Phi_n d\tau = -GM_\odot \mu \left\langle \frac{1}{r} \right\rangle_n == -GM_\odot \mu \frac{1}{\tilde{r}_n}. \tag{23}$$

In equation (23) $\tilde{r}_n^{-1}$ is reciprocal of $\tilde{r}_n$ such as the relation $\dfrac{1}{\tilde{r}_n} = \left\langle \dfrac{1}{r} \right\rangle_n$ holds.

Equating (22) and (23), one obtains the relation

$$\tilde{r}_n = \frac{4D^2}{GM_\odot} n^2 . \tag{24}$$

From this, we get final relation

$$\sqrt{\tilde{r}_n} = \frac{2D}{\sqrt{GM_\odot}} n . \qquad n = (1,2,3,\cdots) \tag{25}$$

Putting $n = m+1$ and

$$\frac{2D}{\sqrt{GM}} = K \tag{26}$$

we come to the final expression

$$\sqrt{R_m} = K(m+1) . \tag{27}$$

Here, $R_m$ replaces the distance $\tilde{r}_m$.

If we suppose that the $n^{th}$ planet has been formed through the mutual gravitating force of the particles of $n^{th}$ band of the diffusing cloud then the distribution law expressed by the relation (27) would become, in that form, the distribution law of the planets of the solar system.

The number $n$ in equation (25) starts from 1, so the number $m$ in equation (27) should be started from zero, 0, which is, however, not allowed. The reason is that as $l = 0,1,2,3,\cdots(n-1)$, the case $n=1$ corresponds to $l=0$, and, from the relation (19), $L_l = d\sqrt{l(l+1)} = 0$, which means that the planet corresponding to the number $n=1$ can not revolve round the Sun. Therefore, the number $m$ must be started from one, that is $m = 1 (n = 2)$.

### 4. Comparison with observation for the planets

In Ref. [4, 5] the distribution law (27) was compared with actual distances of the planets. The Table I shows the comparison of the theoretically calculated $\sqrt{R_m}$ with observation. As we see in Table I, in the inner group of planets one assigns the number $m = 2,3,4,5$ to Mercury, Venus, Earth, and Mars, and in the outer group $m = 1,2,3,4,5$ to Jupiter, Saturn, Uranus, Neptune and Pluto, respectively. Then the relation (27) is splendidly satisfied within the accuracy of 3%, provided that $K_{in} = 0.21$ for the inner group and $K_{out} = 1.08$ for the outer group of planets.

In Table I, the column $K$ is an average of the column $\sqrt{R_m}/(m+1)$ obtained from the observation.

For the relation (27) one can write a relation

$$\frac{\sqrt{R_m}}{\sqrt{R_{m'}}} = \frac{m+1}{m'+1}, \qquad (28)$$

where $m$ and $m'$ are the number of orbits of any planets belonging to the same group. In this formula (28), the constant $K$ is not explicitly appeared, so the formula (28) can be thought to be more general compared to the formula (27).

The Table II shows the ratio $\sqrt{R_m}/\sqrt{R_{m'}}$ from the observation and the ratio of corresponding $m+1$ and $m'+1$.

Then the Earth is taken as a fiducial planet $m' = 4$ for the inner group and Uranus $m' = 3$ for the outer group. The consistency between the ratios $\sqrt{R_m}/\sqrt{R_{m'}}$ and $(m+1)/(m'+1)$ is excellent. In all the calculation of Table I, II distance is in astronomical unit, 1au=$1.496 \times 10^{13} cm$.

Table I. Comparison of theoretical $\sqrt{R_m}$ with observation for the planets

|  | planet | $m$ | $\sqrt{R_m}$ observation | $\sqrt{R_m}$ theory | $\sqrt{R_m}/(m+1)$ (observation) | $K$ (observation) |
|---|---|---|---|---|---|---|
| Inner planets | ? | 1 | --- | 0.42 | - | 0.21 |
|  | Mercury | 2 | 0.62 | 0.63 | 0.21 |  |
|  | Venus | 3 | 0.85 | 0.84 | 0.21 |  |
|  | Earth | 4 | 1 | 1.05 | 0.20 |  |
|  | Mars | 5 | 1.23 | 1.26 | 0.21 |  |
|  | Asteroid | 6 |  | 1.47 |  |  |
| Outer planets | Jupiter | 1 | 2.28 | 2.16 | 1.14 | 1.08 |
|  | Saturn | 2 | 3.09 | 3.24 | 1.03 |  |
|  | Uranus | 3 | 4.38 | 4.32 | 1.10 |  |
|  | Neptune | 4 | 5.48 | 5.40 | 1.10 |  |
|  | (Pluto) | 5 | 6.29 | 6.48 | 1.05 |  |

A most eminent property of the new distribution law (27) is in contrast to the previous formulae, to predict a possibility of existence of an unknown planet between the Sun and Mercury. In the Table I and II it is a planet corresponding to number $m=1$. However, because the orbit $m=1$ is very close to the sun, the planet does not to be created due to strong thermal radiation from the protosun, and even if it would have been formed, it, having lost the energy on account of the friction between dense, hot gas surrounding the protosun and the dense gaseous matter of the first band $(m=1)$, had been thought to be merged into the protosun. It is interesting to expect and to look for the unknown planet with the number $m=1$, though. According to the relation (27), the average distance $R_1$ and expected orbital period are

$$\left. \begin{array}{l} R_1 = 0.17\,au, (\approx 2.5 \times 10^7\,km) \\ T_1 = 0.07\,year, (\approx 25\,days) \end{array} \right\} \quad (29)$$

The period $T_1$ calculated according to the Kepler's third law $R^3/T^2 = 1$ ($R$ in unit $R_\oplus = 1$, and $T$ in unit $T_\oplus = 1$). In the 19$^{th}$ century and even in the 20$^{th}$ century, there has been a number of manifestations on the founding of a hypothetical planet in the inside of Mercury's orbit. The hypothetical planet, however, is known to be not discovered yet.

Tale II. Comparison of ratio $\sqrt{R_m}/\sqrt{R_{m'}}$ from the observation with corresponding ratio $(m+1)/(m'+1)$

|  | Planet | $m$ | $\sqrt{R_m}/\sqrt{R_{m'}}$ observation | $(m+1)/(m'+1)$ |
|---|---|---|---|---|
| Inner Planets ($m' = 4$) |  | 1 | --- | 0.4 |
|  | Mercury | 2 | 0.62 | 0.6 |
|  | Venus | 3 | 0.85 | 0.8 |
|  | Earth | 4 | 1 | 1 |
|  | Mars | 5 | 1.23 | 1.2 |
|  | Asteroid | 6 | --- | 1.4 |
| Outer Planets ($m' = 3$) | Jupiter | 1 | 0.52 | 0.5 |
|  | Saturn | 2 | 0.71 | 0.75 |
|  | Uranus | 3 | 1 | 1 |
|  | Neptune | 4 | 1.23 | 1.25 |

## 5. Satellites of planets in the solar system

The formation of the satellites of every planet in the solar system is considered to begin after the protoplanets have been basically built up. The constituent of the medium surrounding the protoplanets have been considerably distinguished from the state preceding the formation of planets in that the primordial nebula round the protosun consisted of mainly molecular gas whereas the surroundings of the protoplanets before the formation of the satellites consisted of a mixture of the molecular gas and dust, the gas constituent making the main portion in weight. Therefore, the diffusion law described in the previous sections would act as it did before for the pure gas. On the other hand, the dust and small condensates were only under gravitational force of the central planet and surrounding medium. Therefore, we will consider here only the band formation of the gas constituent due to the diffusion law controlled by the relations (18), (21) and (27).

(1). Satellites of Jupiter

In Table III, the data on the satellites of Jupiter are presented.

The satellites of Jupiter can be divided into three groups according to inclination of the orbit plane. The distance law (27) can only find for the first group which has the inclination angle of about 0° with respect to the ecliptic plane. The second group from Leda to Elara has a different inclination of about 30° while the third group about 150°, which means that the second and the third groups would have had the quite different histories for their formation.

We will only consider the first group of satellites here, in connection with the distance law (27). When the number of satellites $m$ is assigned as in the Table III, the root of the distance $R_m$ divided by $(m+1)$ is almost constant, which is equal to $K = 0.8$ in average. (The distances of satellites are estimated in unit of radius of the central planet, Jupiter ($R_{Jupiter} = 71492 km$)). The nearest satellites, Metis and Adrasthea have almost the same distance from Jupiter, so they, for example, must be a result of fragmentation of a preexisted satellite by impact of a body such as asteroid.

Table III.　Satellites of Jupiter ($R_{Jupiter} = 71942 km$)

| $m$ | Satellite | $r(10^3 km)$ | Inclination | Mass($10^{20} kg$) | $r/R_{Jupiter} = \bar{r}$ | $\sqrt{\bar{r}}$ | $\dfrac{\sqrt{\bar{r}}}{m+1} = K$ | $\bar{K}$ | Year of discovery |
|---|---|---|---|---|---|---|---|---|---|
|  | Metis | 127.960 | (0) | $9 \times 10^{-4}$ | 1.79 |  |  |  | 1979 |
|  | Adrastea | 128.980 | (0) | $1 \times 10^{-4}$ | 1.80 |  |  |  | 1979 |
| 1 | Amalthea | 181.300 | 0°.4 | $8 \times 10^{-2}$ | 2.54 | 1.60 | 0.80 |  | 1892 |
|  | Thebe | 221.900 | 0°.8 | $1.4 \times 10^{-3}$ | 3.17 |  |  |  | 1979 |
| 2 | Io | 421.600 | 0°.04 | 893.3 | 5.89 | 2.43 | 0.81 |  | 1610 |
| 3 | Europa | 670.900 | 0°.47 | 479.7 | 9.38 | 3.06 | 0.77 |  |  |
| 4 | Ganimede | 1070.000 | 0°.19 | 1482 | 15.00 | 3.87 | 0.77 | 0.80 |  |
| 5 | Callisto | 1883.000 | 0°.28 | 1076 | 26.48 | 5.14 | 0.86 |  |  |
|  | Leda | 11094.000 | 27° | $4 \times 10^{-4}$ | 155.2 |  |  |  | 1974 |
|  | Himalia | 11480.000 | 28° | $8 \times 10^{-2}$ | 160.6 |  |  |  | 1904 |
|  | Listea | 11720.000 | 29° | $6 \times 10^{-4}$ | 163.7 |  |  |  | 1938 |
|  | Elara | 11737.000 | 28° | $6 \times 10^{-3}$ | 164.2 |  |  |  | 1905 |
|  | Ananke | 20200.000 | 147° | $4 \times 10^{-4}$ | 282.6 |  |  |  | 1951 |
|  | Carme | 22600.000 | 163° | $9 \times 10^{-4}$ | 316.1 |  |  |  | 1938 |
|  | Pasiphae | 23500.000 | 147° | $1.6 \times 10^{-3}$ | 328.7 |  |  |  | 1908 |
|  | Sinope | 23700.000 | 153° | $6 \times 10^{-4}$ | 331.5 |  |  |  | 1914 |

　　Therefore in exploring the distance distribution of the satellites, Metis and Adrasthea should be treated as one body. They do not belong to the first group. In fact, a possible smallest number $m = 0$ yields $K(m+1) = 0.8$ while the actual distance yields $\sqrt{R_0} = \sqrt{1.79} \approx 1.38$ which is in discordance with the prediction 0.8.

The distance law (27) is, therefore, possible to apply only to Amalthea ($m=1$), Io ($m=2$), Europa ($m=3$), Ganimede ($m=4$) and Callisto ($m=5$). Forth satellite, Thebe, has a millionth as small mass as Io and Callisto, so though it belongs to the first group (inclination is about $0°.8$ largest of the first group), its formation must be searched in another way.

The second and the third group of the satellites have quite different inclination of the orbital plane and their masses are one millionth or less as small as large satellites, so they would have had also quite different histories of the formation.

The most sensible feature in Table III is that the total mass of the five satellites which follow the distance law (27) is amounted to more than 99.9% of the total mass of all Jupiter's satellites. This fact testifies that the diffusion process in primordial cloud round the planet (Jupiter) had been a crucial one in the formation of the satellites and in the determination of their spatial distribution.

(2). Satellites of Saturn

Ref. [5, 7] had Janus, Mimas, Enceladus, Tethys, Dione and Rhea followed the distance law (27), according to which they have the number $m=4, 5, 6, 7, 8$ and $9$, respectively, and constant $K=0.29$. After the 1960's, ten satellites had been anew discovered. However, the nearest satellites, Pan, Atlas, Prometheus, Pandora and Epimetheus have from one millionth to one thousandth as small mass as large satellites, Mimas and so on (Table IV). It is, therefore, natural to consider their formation in a different way not to follow the path of the formation of the massive satellites. Epimetheus and Janus have almost the same distance from the central planet (151.42 and $151.47 \times 10^3 km$) so they can be thought of as two parts of one body. Furthermore, Tethys, Calipso and Telesto have also almost the same distances (294.66, 294.66, and $294.67 \times 10^3 km$) and the masses of Calipso and Telesto have a hundred thousandth as small as Tethys, so Calipso and Telesto can be thought of as parts of Tethys. Actually, Telesto and Calipso lie at Lagrangean points L₄ and L₅ in the orbit of the much larger Tethys. Dione and Helene are far away from the central planet in the same distance as well and the mass of Helene is a hundred thousandth as small as the mass of Dione. Helene precedes Dione, keeping 60° ahead of Dione, which means that Helene lies at Lagrangean point L4.

Table IV. Satellites of the Saturn ($R_{Saturn} = 60268$)km

| m | Satellite | $r(10^3 km)$ | Inclination | Mass($10^{20}kg$) | $r/R_{Saturn} = \bar{r}$ | $\sqrt{\bar{r}}$ | $\dfrac{\sqrt{\bar{r}}}{m+1} = K$ | $\bar{K}$ | Year of discovery |
|---|---|---|---|---|---|---|---|---|---|
|   | Pan        | 133.58 | ----   | 4.2x10$^{-6}$ | 2.22 | 1.50 |      |      | 1990 |
|   | Atlas      | 137.64 | (0)    | 1.6x10$^{-4}$ | 2.28 |      |      |      | 1980 |
|   | Prometheus | 139.35 | (0)    | 1.4x10$^{-3}$ | 2.31 |      |      |      | 1980 |
|   | Pandora    | 147.1  | (0)    | 1.3x10$^{-3}$ | 2.35 |      |      |      | 1980 |
|   | Epimetheus | 151.42 | 0°.34  | 5.6x10$^{-3}$ | 2.51 |      |      |      | 1966 |
| 4 | Janus      | 151.47 | 0°.14  | 2.0x10$^{-2}$ | 2.51 | 1.58 | 0.32 |      | 1966 |
| 5 | Mimas      | 185.52 | 1°.5   | 0.37          | 3.08 | 1.76 | 0.29 |      | 1789 |
| 6 | Enceladus  | 238.02 | 0°.02  | 0.65          | 3.95 | 1.99 | 0.28 |      | 1789 |
| 7 | Tethys     | 294.66 | 1.°09  | 6.17          | 4.89 | 2.21 | 0.28 | 0.29 | 1684 |
|   | Calypso    | 294.66 | (0)    | 4x10$^{-5}$   | 4.89 |      |      |      | 1980 |
|   | Telesto    | 294.67 | (0)    | 6x10$^{-5}$   | 4.89 |      |      |      | 1980 |
| 8 | Dione      | 377.4  | 0°.02  | 10.8          | 6.26 | 2.50 | 0.28 |      | 1684 |
|   | Helene     | 377.4  | 0°.2   | 1.6x10$^{-4}$ | 6.26 |      |      |      | 1980 |
| 9 | Rhea       | 527.04 | 0°.35  | 23.1          | 8.74 | 2.96 | 0.29 |      | 1672 |
| 15| Titan      | 1221.85| 0°.33  | 1345.5        | 20.3 | 4.50 | 0.28 |      | 1655 |
| 16| Hyperon    | 1481.1 | 0°.43  | 0.28          | 24.6 | 4.96 | 0.29 |      | 1848 |
| 26| Iapetus    | 3561.3 | 14°.72 | 15.9          | 59.1 | 7.69 | 0.28 |      | 1671 |
| 49| Phoebe     | 12952  | 175°.  | 0.1           | 215.0| 14.6 | 0.29 |      | 1898 |

Outermost satellites Iapetus and Phebe have inclinations 14°.72 and 175°, respectively, to the ecliptic plane, so their formation should be considered in a different way. Titan that is most massive out of the Saturn's satellites is twice as far away as Rhea from the central planet, so one can assign number $m=15$ to Titan and number $m=16$ to Hyperion. The why the gap between Rhea and Titan is so wide is likely to be explained by the massiveness of Titan; the matter of the inner part and the outer part of Titan's orbit except Hyperion corresponding to number $10 \leq m < 14$ and $17 < m$ might have been gravitated and merged by the massive Titan.

In the case of Saturn, the satellites that follow the distance law (27) make up also a majority of total mass of the satellites, which substantiates the fact that the crucial mechanism of the formation of the satellites and of the distance distribution would have been the diffusion process of the primordial gas cloud.

(3) Satellites of Uranus

In the 1960's, the five satellites of Uranus, Miranda, Ariel, Umbriel, Titania and Oberon had been known. At present, the satellites are amounted to fifteen, and the newly discovered satellites are all distributed inside of Miranda's orbit (see Table V). Ref [5, 7] had assigned number $m=4,5,6,8$ and 9 to Miranda, Ariel, Umbriel, Titania and Oberon, respectively. The constant $K$ is 0.46 (Table V). After the 1960's, up to now on in addition to them, ten satellites had anew discovered and they all have one part of thousand or less as small mass as old-known heavy satellites.

Among the newly discovered satellites, Cordelia and Puck can be assigned the number $m=2$ and 3, respectively. Their rooted distances from central planet, Uranus, are 1.39 and 1.83, respectively, whereas Ref. [5, 7] had predicted their expected rooted distance as 1.39 and 1.84, which are well in accord with the actual rooted distance. The anew discovered satellites between Cordelia and Puck can not be assigned by any number $m$. Their total mass makes up a minority of the total mass of the massive satellites.

In Table V, the order number $m=7$ is missed, which might be interpreted, for example, by merging of a body corresponding to $m=7$ into the body with number $m=8$. Actually, the mass of Titania ($m=8$) is 3 times as large as Umbriel ($m=6$), which might be a result of the incorporating of two bodies. The merging of two bodies, in turn, could be caused by a collision of a third body from outside.

Table V. Satellites of Uranus ($R_{Uranus}$ =25559km)

| $m$ | Satellite | $r(10^3 km)$ | Inclination | Mass($10^{20}kg$) | $r/R_{Uranus} = \bar{r}$ | $\sqrt{\bar{r}}$ | $\frac{\sqrt{\bar{r}}}{m+1} = K$ | $\bar{K}$ | Year of Discovery |
|---|---|---|---|---|---|---|---|---|---|
| 2 | Cordelia | 49.75 | (0°.14) | 1.7x10⁻⁴ | 1.94 | 1.39 | 0.46 | | 1986 |
|   | Ophelia | 53.77 | (0°.09) | 2.6x10⁻⁴ | 2.11 | | | | 1986 |
|   | Bianca | 59.16 | (0°.16) | 7x10⁻⁴ | 2.32 | | | | 1986 |
|   | Cressida | 61.77 | (0°.04) | 2.6x10⁻³ | 2.42 | 1.55 | | | 1986 |
|   | Desdemona | 62.65 | (0°.16) | 1.7x10⁻³ | 2.46 | | | | 1986 |
|   | Juliet | 64.63 | (0°.06) | 4.3x10⁻³ | 2.53 | | | | 1986 |
|   | Portia | 66.1 | (0°.09) | 1x10⁻² | 2.60 | | | | 1986 |
|   | Rosalind | 69.63 | (0°.28) | 1.5x10⁻³ | 2.74 | | | 0.46 | 1986 |
|   | Belinda | 75.25 | (0°.03) | 2.5x10⁻³ | 2.94 | | | | 1986 |
| 3 | Puck | 86.00 | (0°.31) | 5x10⁻³ | 3.36 | 1.83 | 0.46 | | 1985 |
| 4 | Miranda | 129.8 | 3°.4 | 0.66 | 5.08 | 2.25 | 0.45 | | 1948 |
| 5 | Ariel | 191.2 | 0°.00 | 13.5 | 7.48 | 2.73 | 0.46 | | 1851 |
| 6 | Umbriel | 266.0 | 0°.00 | 11.7 | 10.41 | 3.23 | 0.46 | | 1851 |
| 8 | Titania | 435.8 | 0°.00 | 35.2 | 17.05 | 4.13 | 0.46 | | 1787 |
| 9 | Oberon | 582.6 | 0°.00 | 30.1 | 22.79 | 4.77 | 0.48 | | 1787 |

(4) Satellites of Neptune

Up to 1960's, the satellites of Neptune had been only known two, Triton and Nereid, so Ref. [5, 7] had not considered whether they followed the distance law (27) or not.

At present, however, the number of satellites of Neptune amounts to eight. The largest is Triton, but its orbit plane is inclined 157° to the ecliptic plane, and next satellite, Nereid has an inclination 29°, so their formation would have not followed the path of the nearer satellites from Naiad to Proteus which have the inclination about 0° to the ecliptic plane. Therefore, the satellites of Neptune are divided also into three groups according to the inclination of the orbital plane. Examination of the distance of the satellites from Naiad to Proteus shows that one can assign the number $m = 4, 5, 6$, and 7 to Naiad, Galatea, Larissa and Proteus, respectively, and constant $K$ is 0.27. Thalassa and Despina have almost the same distances as Naiad from the central planet, Neptune, so they are likely to be fragmented from a satellite by impact of a third body. In the case of Neptune's satellites, the distance law (27) satisfies also for the heavy satellites of first group.

Table VI. Satellites of Neptune ($R_{Neptune}$ =24764 km)

| $m$ | Satellite | $r(10^3 km)$ | Inclination | Mass($10^{20} kg$) | $r/R_{Neptune} = \bar{r}$ | $\sqrt{\bar{r}}$ | $\dfrac{\sqrt{\bar{r}}}{m+1} = K$ | $\bar{K}$ | Year of discovery |
|---|---|---|---|---|---|---|---|---|---|
| 4 | Naiad | 48.2 | (0) | $1.4 \times 10^{-3}$ | 1.95 | 1.39 | 0.28 | | 1989 |
| | Thalassa | 50.0 | (4°.5) | $4 \times 10^{-3}$ | 2.02 | 1.43 | | | 1989 |
| | Despina | 52.5 | (0) | $2.1 \times 10^{-3}$ | 2.12 | 1.45 | | | 1989 |
| 5 | Galatea | 62.0 | (0) | $3.1 \times 10^{-2}$ | 2.50 | 1.58 | 0.26 | | 1989 |
| 6 | Larissa | 73.6 | (0) | $6 \times 10^{-2}$ | 2.97 | 1.73 | 0.25 | 0.27 | 1989 |
| 7 | Proteus | 117.6 | (0) | 0.6 | 4.75 | 2.18 | 0.27 | | 1989 |
| | Triton | 354.59 | (157°) | 214 | 14.3 | 3.78 | | | 1846 |
| | Nereid | 5588.6 | (29°) | 0.31 | 226 | 15.0 | | | 1949 |

## 6. Discussion and conclusion

In this work, we have mainly focused on the role of the diffusional wave character in the formation of the distance configuration of the solar system. The distance law (27) applied to the planets does excellently agree with observation. The separation into the inner and outer parts of the planets in the solar system is also accounted for by the diffusion character provided that the constant $K$ which reflects the different physical conditions of the diffusion medium of the inner and outer parts have values 0.21 and 1.08 for inner and outer parts of the protonebula surrounding the protosun, respectively.

In this paper, the satellites of the Earth and Mars have not been considered. Actually, the terrestrial planets are poor in satellites. The Moon of the Earth is considered to be formed through an impact of an asteroid, and two satellites, Phobos and Deimos are believed to be asteroids captured from the asteroid belt. Mercury and Venus are absent of satellites. In the inner group of planets, therefore the band formation due to diffusional wave character in the protonebula surrounding the planets seems to be not so sensible as in the satellites of Jovian planets. Instead, the outer group of planets has rich satellite system and, recently it is found every Jovian planet to have a ring inside of the satellite system. In the satellite system of Jovian planets, the agreement of the distance law (27) with observation is sensible only for the massive satellites. This fact substantiates that the radical process in the formation of the satellites having the majority of mass had been the diffusional wave character in the gaseous nebula surrounded the protoplanets. The satellites which have one thousandth or less as small mass as the massive one are placed mainly in the inner part of the satellite system and are contiguous to the ring system. Therefore, the low-mass satellites could be found its mechanism of the formation in a way somewhat common to the ring.

The fact that the satellite system had also governed by the diffusion law says that the nebula up to the period of the formation of satellite system had been in almost gaseous, homogeneous state and in temperature of a few hundreds degree.

Initial ingredients of the solar nebula consisted of hydrogen and helium gas (98%), hydrogen compounds (1.4%), rock (0.4%) and metal (0.2%) molecules. Hydrogen and helium gas did not condense and vast majority of the nebula remained gaseous at initial time of the formation of the solar system. Hydrogen compounds condense at lower temperature than about 200k. Rocky minerals condense at 500-1300K and metals at 1000-1600K.

At first, the inner part of nebula was heated to about two thousands degree due to irradiation from the protosun and, therefore, all the ingredients of inner part were in gaseous state. The more far away from the protosun, the lower the temperature of the nebula and, in a few au's (near current Jupiter's orbit) from the protosun, temperature descended down to a few hundred degree. Therefore, in the outer part of nebula the hydrogen compounds, including rock and metals, began to condense. Nevertheless, the remaining hydrogen and helium gas in outer part of the nebula made up 98% 0f total mass of that part and, therefore, the diffusion law represented by Eq.(15) operated for the gaseous ingredients hydrogen and helium. Instead, in the inner part of nebula, the diffusion law could be in operation for all the ingredients being in gaseous state in so far as the temperature remained above two thousands degree. Therefore, the inner part of the nebular could develop a band configuration following Eq.(15).

With the descending of temperature below about a thousand degree in the inner part of nebula, the rocky and metallic molecules began to condense. Thus solid particles in the inner part of nebula were made only of rock and metals and these particles became the seeds for further condensation. Inside of the bands, temperature was lower than in out, and, therefore, the solidification (condensation) of the rocky and metallic materials was more effective there. On the other hand, the gaseous hydrogen, helium and hydrogen compounds could not be attracted efficiently into the accretion center because of their lightness, and, furthermore, irradiation and wind from central protosun rendered the gas compounds to escape outwards. Therefore, the terrestrial planets became dense. The condensation and the sedimentation (accretion) of the rocky and metallic materials to more and more large particles came to formation of the planetesimals and they grew eventually into planets.

On the other hand, in the outer part of the nebula, rock, metals and hydrogen compounds condensed to solid particles and they grew into planetesimals. Here hydrogen compounds were three times as many as rock and metals, so the sedimented lumps of icy materials of hydrogen compounds, rock and metals were much larger than the case of the inner part. Therefore, the growth of icy planetesimal can not be the whole story of Jovian planet formation because the Jovian planets themselves are not made mostly of ice. Instead, these planets are fraught with hydrogen and helium today. This means that these planets formed as gravity drew gas around large, icy planetesimals.

The initial physical state of the primordial solar nebula which determines the distance distribution of the planets and the satellites could be theoretically predicted from the condition of interstellar molecular clouds where stars and planetary systems are forming. The theoretical study on the constant $K$ will be followed in detail in next work.